# Rolling spinners on the water surface

Jean-Baptiste Gorce,[1] Konstantin Y. Bliokh,[2] Hua Xia,[1] Nicolas Francois,[1] Horst Punzmann,[1] Michael Shats[1*]

[1]Research School of Physics, The Australian National University, Canberra ACT 2601, Australia
[2]Theoretical Quantum Physics Laboratory, RIKEN Cluster for Pioneering Research, Wako-shi, Saitama 351-0198, Japan
*Corresponding author. Email: Michael.Shats@anu.edu.au

**Abstract**:
Angular momentum of spinning bodies leads to their remarkable interactions with fields, waves, fluids, and solids. Orbiting celestial bodies, balls in sports, liquid droplets above a hot plate, nanoparticles in optical fields, and spinning quantum particles exhibit nontrivial rotational dynamics. Here, we report self-guided propulsion of magnetic fast-spinning particles on a liquid surface in the presence of a solid boundary. Above some critical spinning frequency (higher rotational Reynolds numbers), such particles generate localized 3D vortices and form composite 'spinner-vortex' quasi-particles with nontrivial, yet robust dynamics. Such spinner-vortices are attracted and dynamically trapped near the boundaries, propagating along the wall of any shape similarly to 'liquid wheels'. The propulsion velocity and the distance to the wall are controlled by the angular velocity of the spinner via the balance between the Magnus and wall-repulsion forces. Our results offer a new type of surface vehicles and provide a powerful tool to manipulate spinning objects in fluids.

**INTRODUCTION**

The dynamics of rotating matter carrying angular momentum is crucial in numerous problems spanning from microscopic quantum to giant astrophysical scales. These problems involve rotating or spinning particles: from quantum elementary particles *(1)* to planets and stars *(2)*, and a variety of vortices: in classical *(3)* and quantum *(4,5)* fluids, as well as in classical *(6,7)* and quantum *(8)* wavefields. Quite often, the angular momenta of spinning particles and vortices become closely related and mutually coupled *(9–11)*. Not surprisingly, the dynamics of spinning particles and vortices share fundamental similarities, such as the Magnus and various Hall effects, i.e., the transverse angular-momentum-induced transport *(12–15)*.

Spinning particles in classical fluids naturally appear under optical manipulation with vortex or circularly-polarized fields *(9,10)*, in externally imposed magnetic fields *(16–18)*, and in sports ballistics *(19,20)*. Furthermore, liquid droplets on hot solid surfaces also behave as spinning liquid particles or 'Leidenfrost wheels' *(21–23)*. Recently it was shown that water surface waves generating circular flows and polarizations *(24)* can efficiently interact with sub-wavelength spinners, which allows efficient manipulation of particles on the water surface *(25)*. Therefore, understanding the dynamics of spinning particles in fluids constitutes a both fundamental and applied problem important for physics and engineering.

Here we report a novel behavior in a known classical system: a spinning particle (spinner) floating on the water surface. We show that above some threshold angular velocity (i.e., above critical Reynolds number), the spinner generates a well-localized 3D vortex around it, and this provides an effective coupled *spinner-vortex* 'quasi-particle'. Most remarkably, such spinner-vortices become attracted and trapped near solid boundaries (where the wall-normal repulsion and Magnus forces balance each other) and roll with constant translational velocities along the boundaries of arbitrary shapes. This provides robust 'liquid wheels' which can be controlled via the angular velocity of the spinners and transported along any desired boundary without touching it. We find simple and robust dependencies between the main parameters of such self-guided propulsion: the angular velocity, the linear velocity, and the distance to the boundary. Our results provide a novel type of the spinner-vortex coupling in classical hydrodynamics and suggest a new type of robust self-navigated transport which can be employed in liquid surface vehicles.

## RESULTS

### The main parameters and phenomenon

We study the motion of fast-spinning particles floating on the water surface in the absence of externally imposed flows but in the presence of fixed solid boundaries. Particles studied here are magnetized disks with radii in the range of $a = (0.5 - 2.5)$ mm and a thickness of $h = 1$ mm. The spinning motion of the particles is generated and supported by an external rotating magnetic field (see the Supplementary Materials), such that the spinning frequency can be varied in the range of $f = \omega/2\pi = (1-50)$ Hz. The corresponding rotational Reynolds numbers are $Re_\omega = 2a^2\omega/\nu = (20-200)$, where $\nu = 10^{-6}$ m$^2$/s is the kinematic viscosity of water. In these conditions, inertial effects must play an important role for the mechanisms of the spinner's propulsion and its interaction with a solid boundary.

Most remarkably, we observed that fast-spinning particles, that float on the water surface, are attracted to solid boundaries and roll along these boundaries of practically any shape with a fixed velocity $V$ at a fixed distance $\delta$ from the boundary. Figure 1 shows examples of the spinner motion around triangular, star-shaped, and irregular 'Australia-shaped' boundaries (see also the Supplementary Movies S1–S3).

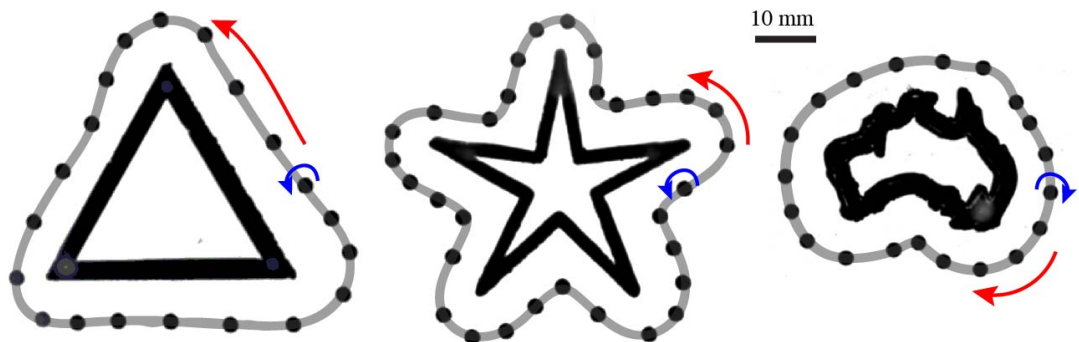

**Fig. 1. Fast-spinning floating particles move along complex-shape boundaries.** The spinner radius is $a = 1$ mm, the spinning frequency is $\omega/2\pi = 12$ Hz. Time between consecutive frames is $\Delta t = 1$ s. Note the reversed angular and linear velocities in the motion along the 'Australia-shaped' boundary. See also the Supplementary Movies S1–S3.

### Spinner-vortex quasi-particle

Spinners interact with other objects in a fluid by creating *vortices* around themselves. Therefore, we first look at the flow created by a spinning disk away from the wall. At low rotation frequency $\omega$ (or in a viscous fluid), the spinner creates a vortex around its axis of rotation. The fluid's azimuthal velocity in the vortex at the distance $\rho$ from the center of rotation is given by $v_\vartheta(r) = a^3\omega/\rho^2$ *(18)*. Indeed, we observe such a vortex with purely azimuthal flow at low

Page 2 of 10

rotational Reynolds numbers $Re_\omega < 20$, as shown in Fig. 2 (A and B). This experiment is performed on the surface of a glycerol solution, whose viscosity is 50 times higher than that of water. In this regime, the spinner remains motionless, and no propulsion is observed.

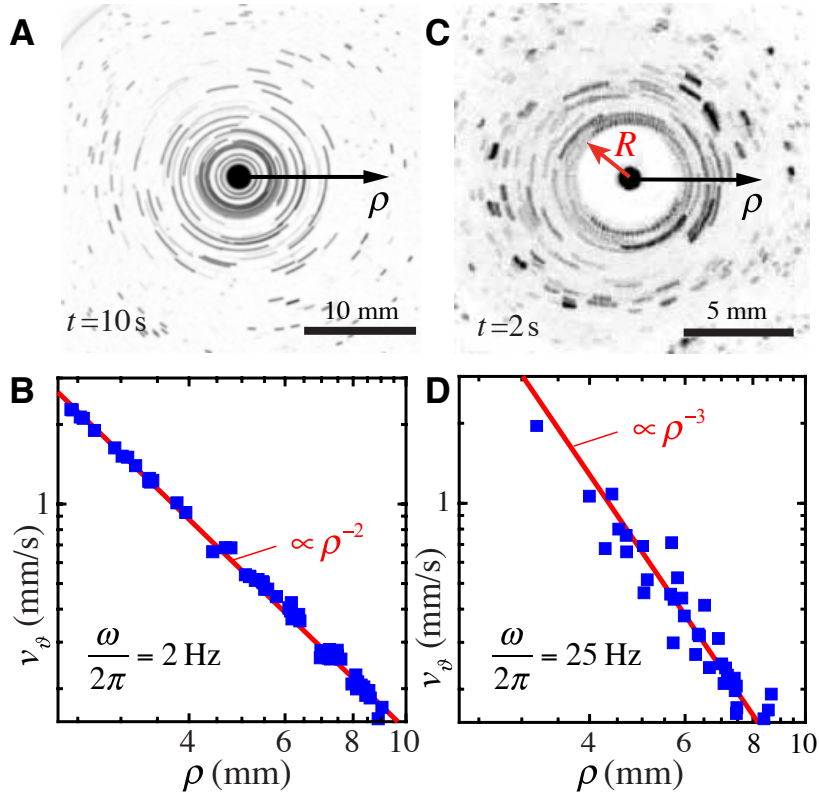

**Fig. 2. Generation of the 'spinner-vortex' quasi-particle.** (**A**) The flow around a spinner with $a = 0.5$ mm and angular velocity $\omega < \omega_0$ ($Re_\omega \simeq 6$ here). Here the spinner floats on the surface of a glycerol-water solution. (**B**) Radial dependence of the azimuthal velocity in the vortex flow from (**A**). (**C**) and (**D**) Same as in (**A**) and (**B**) but for the spinner with $\omega > \omega_0$ on a water surface ($Re_\omega \simeq 79$ here). A localized 3D vortex characterized by a depletion zone with the radius $R \simeq 2.5$ mm develops.

As the frequency is increased above some critical value, $\omega > \omega_0$, which corresponds to the rotational Reynolds numbers $Re_\omega > Re_{\omega 0} = (22 \pm 2)$, the vortex around the spinner changes qualitatively. Namely, fluid particles are expelled from the vicinity of the spinner, forming a finite-size depletion zone whose radius $R$ is independent of the angular velocity $\omega$ for a given spinner size (Fig. 2C). Such a depletion zone was previously reported with regard to spinners on the liquid surface in Ref. *(16)*, where it was referred to as the high-pressure region. The radial profile of the azimuthal velocity of the fluid in this regime is different from that at lower angular velocity $\omega$ (lower $Re_\omega$): outside the depletion zone ($\rho > R$) the velocity decreases as $v_\vartheta \propto \rho^{-3}$ (Fig. 2D). We emphasize that this well-localized vortex is a 3D object with essential vertical flow components; its 2D flow does not preserve the number of fluid particles. In what follows, we consider a fast-spinning particle surrounded by the localized vortex as a single *spinner-vortex* quasi-particle.

**Dynamical instability and attraction to the boundary**

Importantly, the motionless state of the spinner-vortex with $\omega > \omega_0$ becomes *unstable*. It undergoes small random excursions from the initial position and then starts moving with exponentially growing linear velocity towards the wall (see the Supplementary Movie S4). When approaching the wall, the spinner-vortex becomes *trapped* at some distance $\delta$, and its velocity $V$ becomes stabilized and directed along the wall. Figure 3A shows an example of such evolution for the spinner originally located in the center of a circular pool and eventually trapped near its boundary. Thus, the only stable regime for such spinner-vortex is the '*liquid-wheel rolling*' along the wall. Note that this motion is in agreement with an intuitive picture of a rolling wheel: the linear velocity is directed as $\mathbf{V} \propto \boldsymbol{\omega} \times \mathbf{n}$, where $\boldsymbol{\omega}$ is the angular velocity vector, while $\mathbf{n}$ is the normal to the boundary directed towards the spinner. Reversing $\boldsymbol{\omega}$ or placing the spinner on the opposite side from the wall reverses the direction of its linear motion.

When the spinner becomes unstable and starts moving within such a circular pool, its trajectory is well described by the parametric fit:

$$r(\theta) = r_w \frac{\exp(b\theta)}{C + \exp(b\theta)}, \qquad (1)$$

where $(r,\theta)$ are the polar coordinates with the origin in the centre of a circular boundary and $(b, r_w, C \gg 1)$ are constant parameters. The measured trajectory of a spinner, together with the parametric fit Eq. 1, are shown in Fig. 3 (B and C). At short times, or small $\theta$, the radius grows exponentially, $r \propto \exp(b\theta)$, which corresponds to a logarithmic spiral. Later this growth saturates at the stationary radius along the wall-guided orbit at $r = r_w$. The temporal evolution of the spinner velocity, $V(t)$, and azimuthal angle, $\theta(t)$, are shown in Fig. 3 (D and E). One can see that the spinner travels with a constant angular velocity $\Omega = d\theta/dt$ along the spiral part of the trajectory (which is a characteristic feature of the logarithmic spiral) and then travels with another constant angular velocity along the circular wall. Remarkably, the way the spinner-vortex approaches the wall is very robust: the ratio of the spinner velocity $V$ along its trajectory to its terminal velocity along the wall, $V_w$, is independent of its frequency $\omega$ (Fig. 3F).

To get some intuition about dynamical laws underlying this motion, we assume that the spinner-vortex moves with a constant angular velocity $\Omega$, $\theta \simeq \Omega t$, and that the azimuthal velocity component makes the main contribution to the velocity $V$ during the spiral motion: $V \simeq r\Omega$. Then, the evolution of the velocity can be written as

$$V(t) = V_w \frac{e^{\beta t}}{C + e^{\beta t}}, \quad V(s) = V_w \left(1 - e^{-\beta s/V_w}\right), \qquad (2)$$

where $V_w = \Omega r_w$, $\beta = \Omega b$, we introduced the coordinate along the spinner trajectory, $s = \int V dt$, and set the initial condition $V(s=0) = 0$. The temporal and coordinate dependences of the velocity, Eq. 2, fit the experimentally measured dependences in Fig. 3 (D and F) very well. Furthermore, the comparison with Fig. 3F confirms that the parameter $\beta/V_w$ is independent of the spinner frequency $\omega$.

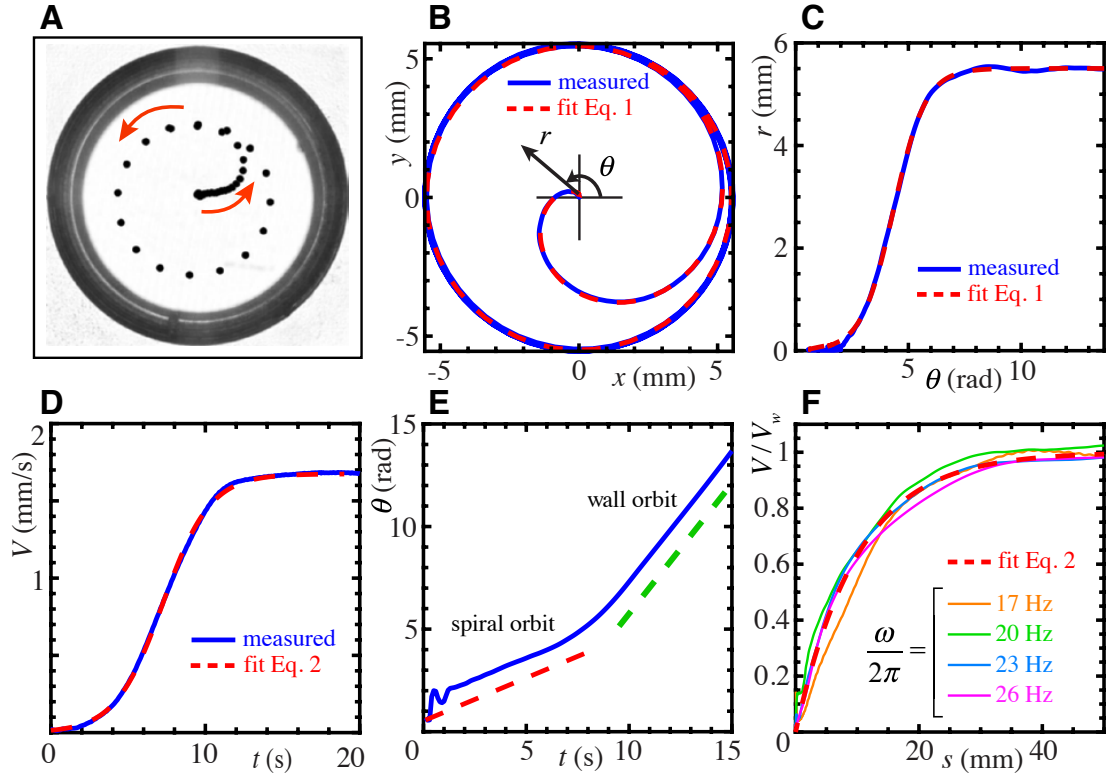

**Fig. 3. Onset of the 'spinner-vortex' mobility and its movement to the boundary.**
(**A**) The 'spinner-vortex' quasi-particle, formed above the $\omega = \omega_0$ threshold, moves from the center of the container towards its circular boundary of 30 mm radius. (**B**) The trajectory of the spinner is a logarithmic spiral which converges to a limit circle near the boundary. The experimentally measured trajectory and the fit given by Eq. 1 are shown. (**C**) The same trajectory presented by the $r(\theta)$ dependence of the polar coordinates of the spinner. (**D** and **E**) Experimentally measured temporal dependence of the spinner-vortex velocity, $V(t)$, and azimuthal angle, $\theta(t)$, along the trajectory in (**B**). (**F**) Velocities of the spinners with different frequencies $\omega/2\pi$, normalized by their velocities near the wall, $V_w$, versus the distance $s$ along their trajectories.

Importantly, the dependence $V(t)$, Eq. 2, suggests the equation of motion along the spinner trajectory as follows:

$$m\frac{dV}{dt} = A(\omega - \omega_0)RV - BV^2 . \qquad (3)$$

Here, $m$ is a mass parameter, $A(\omega - \omega_0)R = \beta m$, $B = A(\omega - \omega_0)R/V_w$, and we took into account that the velocity of the spinner is proportional to its frequency: $V \propto V_w \propto (\omega - \omega_0)R$ (see below). These relations also mean that $\beta \propto (\omega - \omega_0)$ (this is also confirmed directly), while $m$ is frequency-independent.

The two terms on the right-hand side of Eq. 3 can be associated with a positive 'propulsion force' and negative drag (friction) force, respectively. The drag force originates from the viscosity of the fluid, i.e., dissipation in the system, while the propulsion force is produced by the external torque acting on the spinner, which supplies the energy into the system. Since the moving spinner-vortex has an asymmetric shape (due to the Magnus effect, curvature of the trajectory, presence of the wall, etc.), the friction is different on the two sides of the spinner-vortex, and this transforms the external torque into the propulsion force (the 'rolling wheel' mechanism). Naturally, this force acts only at $\omega > \omega_0$, and it grows linearly with $\omega$. Equation 3 describes both the exponential grows of the velocity at the early stage of the spinner-vortex motion, as well as the stabilized motion near the boundary. In the latter case, the two forces balance each other resulting in the uniform motion with $V = V_w$.

**The rolling-wheel motion along the boundary**

To investigate the main features of the uniform 'rolling wheel' motion along the wall, we performed a number of measurements for spinners moving along a circular boundary shown in Fig. 4A. The stable motion of the spinner-vortex along the wall suggests that there is a balance of forces normal to the wall (Fig. 4B) (we neglect the centrifugal force due to the curvature of the orbit). One of these forces is the repulsion force $\mathbf{F}_R \parallel \mathbf{n}$ pushing the spinner away from the wall, while another one is the *Magnus force* pushing the spinner towards the wall. The Magnus force has the form $\mathbf{F}_M = \tilde{m}\boldsymbol{\omega} \times \mathbf{V} \parallel -\mathbf{n}$ *(12,13)*, where $\tilde{m}$ is a parameter with the dimension of mass. From here on we consider the velocity of the spinner near the wall omitting the subscript "w". The fluid streamlines around a spinner moving along the wall, shown in Fig. 4B, are reminiscent of those giving rise to the Magnus (lift) force around spinning balls in sports *(20)*. The spinner-vortex is trapped near the wall because the repulsion force decreases with the distance from the wall, while the Magnus force remains approximately constant. Furthermore, as the angular velocity ω increases, the Magnus force grows and the spinner is pushed closer to the wall, i.e., the distance $\delta$ decreases. This corresponds exactly to the behavior observed in our experiments.

To quantify the relations between the main parameters of the 'rolling wheel' motion, $\omega$, $V$, and $\delta$, we performed a series of measurements to determine these parameters for spinners with different radii $a$ and angular velocities $\omega$. The results shown in Fig. 4 (C and D) reveal rather simple and robust dependencies between these quantities:

$$V = \alpha R(\omega - \omega_0), \quad V = V_0 \exp\left(-\frac{\delta - R}{L}\right). \tag{4}$$

Here, $\alpha$, $V_0$, and $L$ are parameters of no dimension, velocity dimension, and length dimension, respectively, and all of these are independent of the spinner size $a$. The linear dependence $V \propto (\omega - \omega_0)$ also follows from Eq. 3, where $\alpha = A/B$. In our experiments, we found that $\alpha \simeq 0.05$ (this is probably related to the critical rotational Reynolds number: $\alpha \simeq 1/Re_{\omega 0}$), $V_0 \simeq 59$ mm/s, and $L \simeq 2.9$ mm. Notably, the linear proportionality of the translational velocity to the angular velocity, the first Eq. 4, agrees with the intuitive picture of the rolling wheel. Since the effective angular velocity of the rolling motion is $(\omega - \omega_0)$ rather than $\omega$, one can expect that the Magnus force should also be modified as $F_M = \tilde{m}(\omega - \omega_0)V$. In turn, the second Eq. 4 reflects the fact that larger angular and linear velocities produce smaller distances $\delta$ (as the Magnus force increases). The particular exponential dependence is apparently related to the form of the repulsion force. From Eqs. 4, and the form of the Magnus force, we can write an effective form of the repulsion force: $F_R(\delta) = F_0 \exp\left[-2(\delta - R)/L\right]$, where $F_0 = \tilde{m}V_0^2/\alpha R$. Note that $(\delta - R)$ is the distance between the spinner-vortex boundary and the wall.

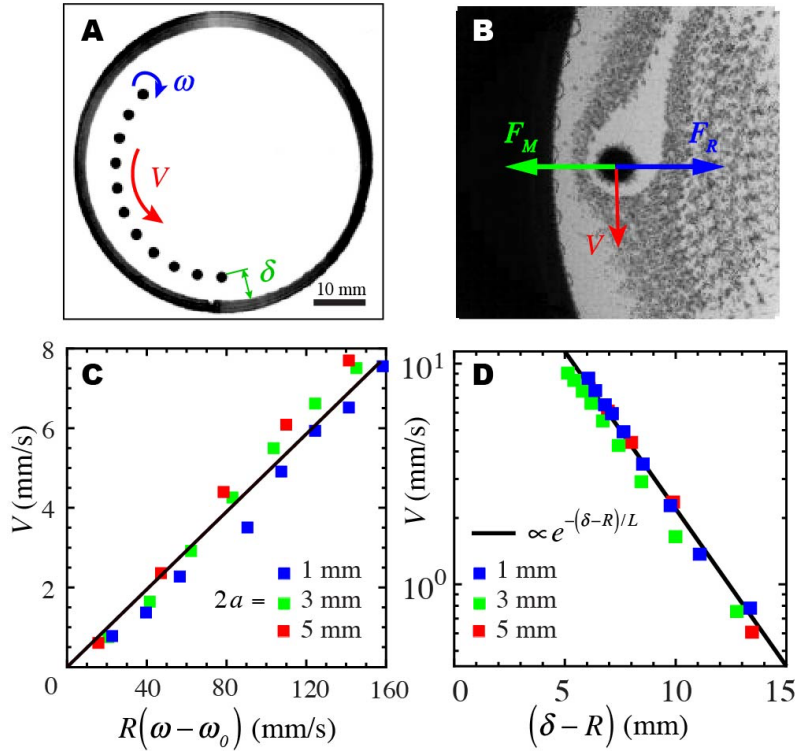

**Fig. 4. 'Liquid-wheel' motion of the spinner-vortex along a circular boundary.** (**A**) Trajectory of a spinner with $a = 0.5$ mm moving inside the circular container. Here its propagational velocity is $V \simeq 12.4$ mm/s at $\delta \simeq 4.3$ mm distance to the wall. (**B**) The water flow around the spinner travelling along the wall. Grey streaks show motion of the tracer particles. The arrows indicate the balance of forces normal to the boundary. (**C**) The linear velocity $V$ of the spinner versus its angular velocity $\omega$ for different spinner sizes $a$. (**D**) The spinner velocity $V$ versus the distance δ between the spinner-vortex and the wall.

A rigorous theoretical calculation of the repulsion force is a rather challenging problem depending on the details of the interaction between a moving localized 3D vortex and the wall. We leave it for further studies and demonstrate the presence of this repulsion force experimentally. For this purpose, we performed a series of measurements using instantaneous switching off the spinning motion, i.e., making $\omega = 0$ at some instant of time $t = t_0$. This switches off the Magnus force, and the spinner starts moving away from the wall under the action of the repulsion force (see the Supplementary Movie S5). Figure 5 shows that the repulsion force pushes the particle away from the border such that its trajectory has the radius of curvature approximately equal to $L \simeq 2.9$ mm, independently of the spinner size $a$ and its velocity $V$ at $t = t_0$. The repulsion force can be estimated as $m_0 V^2 / L$, where $m_0$ is the spinner mass. By equating it to the Magnus force at $t < t_0$, $F_M = \tilde{m}(\omega - \omega_0)V = \tilde{m}V^2/\alpha R$, we find that $\tilde{m} = m_0 \alpha L / R$. This can be used only as a rough estimate because switching off the spinner destroys the localized vortex around it, so that the parameters of this object (mass, size, etc.) can change considerably.

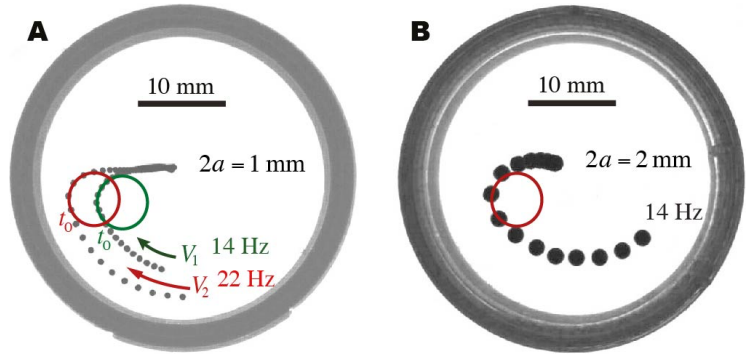

**Fig. 5. Observation of the repulsion force from the wall.** (A) Trajectories of two spinners ($a = 0.5$ mm, $\omega/2\pi = 22$ Hz and $\omega/2\pi = 14$ Hz) travelling along stationary wall orbits. At the instant of time $t = t_0$, the torque supporting the angular velocity of the spinner, $\omega$, is switched off. As a result, the angular velocity and the Magnus force vanish, and the spinners are deflected by the repulsion force (Fig. 4B) with a characteristic curvature radius $L \simeq 2.9$ mm, which is independent of $\omega$ (see the red and green circles of the same radius $L$). (B) A trajectory of the larger spinner with $a = 1$ mm and $\omega/2\pi = 22$ Hz shows the same scattering radius by the wall repulsion force as in (A).

## DISCUSSION

To summarize, we have reported a novel deterministic behavior of fast-spinning particles on the liquid surface. Above some critical rotational Reynolds number, the spinner is dynamically trapped and propels at a fixed distance from a boundary following its arbitrary shape. This phenomenon is accompanied by the generation of a localized 3D vortex around the spinner and is described by the balance of the Magnus and wall-repulsion forces. The effect reported here is very robust and can be controlled by a single parameter: the angular velocity of the spinner. Note also that it does not rely on the surface-specific effects: the buoyancy and the surface tension just balance the gravitational force acting on a spinner. Thus, forces discussed here should also act on any spinning object in the presence of walls.

Our results offer exciting possibilities of developing self-navigating water surface vehicles, with various applications in laboratory and marine robotics, manipulation and sorting of spinning or magnetic particles, mixing of chemical or biological substances, etc. In this work, we have presented detailed experimental observations and measurements, together with their phenomenological analysis. An accurate theoretical description of the observed phenomena is a rather challenging problem for future studies. Indeed, even a toy mechanical model of the spinner-vortex behavior must describe a non-conservative system with gain (via external torque) and loss (via friction) of energy, as well as with coupling between rotational and translational degrees of freedom. Hydrodynamic analysis of the problem is even more challenging because the flows around spinning bodies have been explored so far only in numerical simulations and only in a limited range of Reynolds numbers *(18,26,27)*. Further understanding of the phenomenon of rolling spinners will require experimental and theoretical studies of the 3D fluid motion within the spinner-vortex structure, which seems to be the key to the observed effect. It would also be interesting to investigate in the future the motion of spinners along modulated walls (e.g. sinusoidal), as well as collective wall-guided motion of multiple spinners.

## MATERIALS AND METHODS

**Rotating magnetic field**

The external magnetic field is produced by four Helmholtz coils. The electric current in the coil pairs is driven by two waveform amplifiers (Accel Instruments TS250-0) which create a rotating horizontal magnetic field at the surface of the liquid (Fig. S1). Opposite coils are connected in series and the current in the pairs is phase-shifted by $\pi/2$. The resulting magnetic field rotates at the frequency $\omega/2\pi$. The rotation direction can be reversed by changing the phase between the magnetic coil pairs from $\pi/2$ to $-\pi/2$. The horizontal magnetic field at the liquid surface is approximately constant: $\Delta B/B < 0.02$. This has been tested using a 2D scanning Hall probe at the coils horizontal mid-plane, corresponding to the liquid level.

**Spinners and boundaries**

Ferromagnetic spinners of different diameters are manufactured by 3D printing a template (PLAflex) consisting of a pattern of cylindrical holes. The holes are loaded with a mixture of polymers (Elite Double 8) and nickel powder ($<150\,\mu m$; Sigma Aldrich), where the nickel powder constitutes 90% of the mixture mass. To ensure that the spinner disk is magnetized along its diameter, the 3D printed template containing the spinners is exposed to a horizontal uniform magnetic field of 2 T for 3 days. The density of the spinners is adjusted to avoid the formation of a meniscus at the spinners edge. In addition, the top surface of the spinner is Teflon-coated to avoid wetting. This is done to eliminate the Cheerios effect [28] which could add to the spinner-wall interaction. When placed on the water surface, the torque on the spinner is imposed by the external rotating magnetic field. The spinner rotation frequency is verified using a high-speed camera.

Boundaries and liquid containers are made of PLA thermoplastic using a 3D Ultimaker 2 + printer. To avoid menisci at the container boundary, a step at the level of 15 mm from the bottom of the container is made and a liquid is filled to that level. In experiments summarized in Figs. 3 and 4, the diameter of the circular container is 60 mm.

**Acknowledgements:** This work was supported by the Australian Research Council Discovery Projects and Linkage Projects funding schemes (DP160100863, DP190100406 and LP160100477). H.X. acknowledges support from the Australian Research Council Future Fellowship (FT140100067). N.F. acknowledges support by the Australian Research Council DECRA award (DE160100742).

**Author contributions:** HX, NF, HP and MS designed the project. JBG conducted experiments. JBG, MS, and KB analyzed and interpreted the results and wrote the paper. All authors discussed and edited the manuscript.